\documentclass[11pt,dvips]{article}

\usepackage{epsfig,times} 
\usepackage{picinpar}
\makeatletter
\renewcommand{\section}{\@startsection{section}{1}{0in}
	{0.4\baselineskip}{0.1\baselineskip}{\Large\bf}}
\renewcommand{\subsection}{\@startsection{subsection}{2}{0in}
	{0.25\baselineskip}{-\baselineskip}{\large\bf}}
\renewcommand{\subsubsection}{\@startsection{subsubsection}{3}{0in}
	{0.1\baselineskip}{-\baselineskip}{\normalsize\bf}}
\makeatother
\begin{document}

%
\makeatletter\newcommand{\ps@icrc}{
\renewcommand{\@oddhead}{\slshape{OG.2.1.09}\hfil}}
\makeatother\thispagestyle{icrc}
%
%

\begin{center}
%
{\LARGE \bf OBSERVATIONS OF GAMMA-RAY EMISSION FROM THE BLAZAR MARKARIAN 421 ABOVE 250 GeV
WITH THE CAT CHERENKOV IMAGING TELESCOPE}
\end{center}

\begin{center}
%
%
{\bf F. Piron$^{1}$, for the C{\small AT} collaboration
}\\
{\it $^{1}$LPNHE, Ecole Polytechnique and IN2P3/CNRS, Palaiseau, France\\
}
\end{center}

\begin{center}
{\large \bf Abstract\\}
\end{center}
\vspace{-0.5ex}
%
%
The $\gamma$-ray emission of the blazar Markarian 421 above
$250\:\mathrm{GeV}$ has been observed by the C{\small AT} Cherenkov
imaging telescope since December, 1996. We report here results on the
source variability up to April, 1998, with emphasis on the 1998 campaign.
For the flaring periods of this year, the energy spectrum was derived
from $330\:\mathrm{GeV}$ up to $5.2\:\mathrm{TeV}$: it is very well
represented by a simple power law,
with a differential spectral index of $2.96 \pm 0.13$.\\
\vspace{1ex}

%
%
\section{Introduction}
\label{alpha}
\begin{figwindow}[1,r,
{\mbox{\epsfig{file=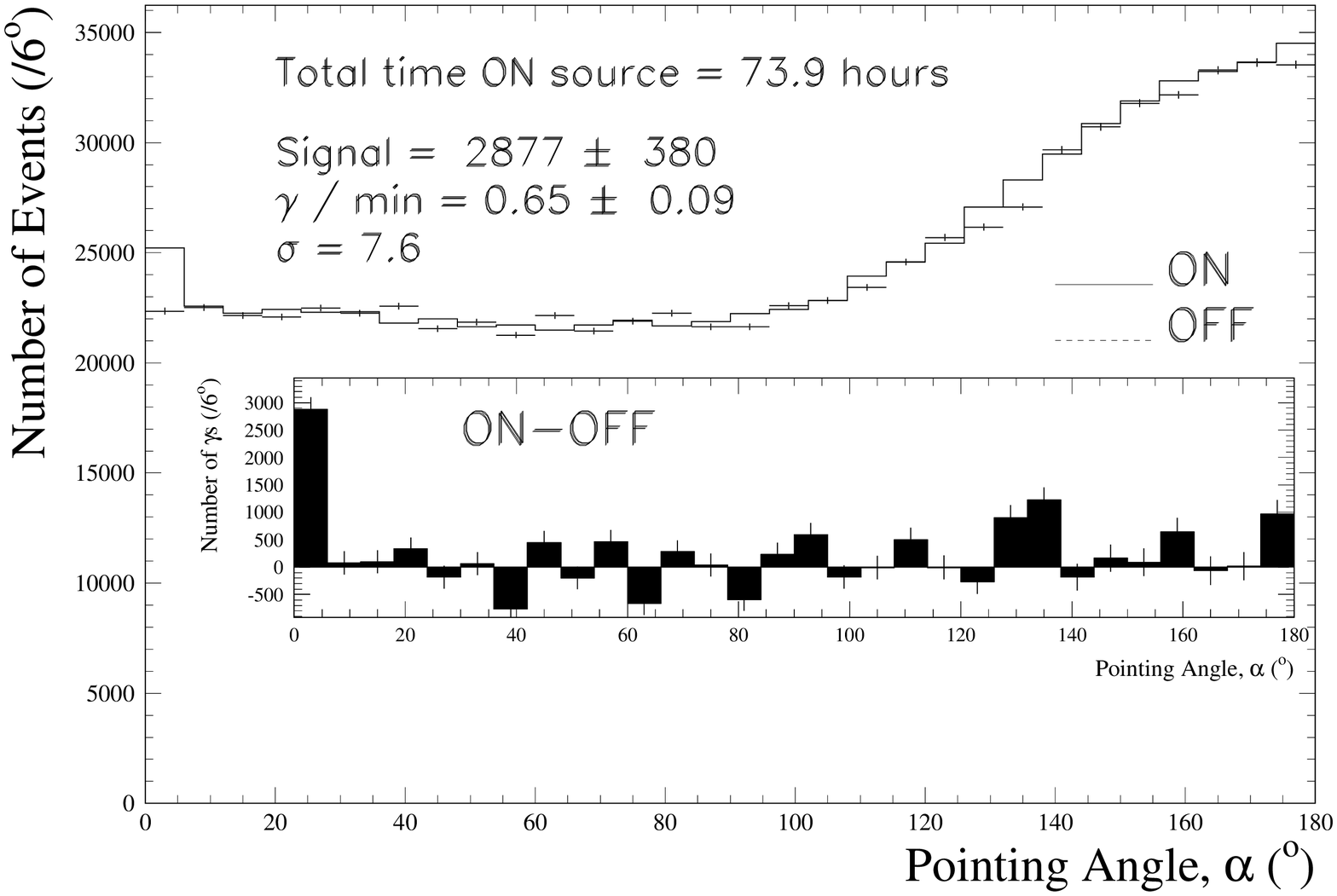,width=4.2in}}},
{\it Distribution of the pointing angle $\alpha$ (see Sect.~\ref{detana.sec} for definition) of Mrk~421
for all data between December 1996 and April 1998. The signal is clearly seen in the first bin.}]
%
%
Markarian~421 (Mrk~421) is one of the three extragalactic sources detected to date from the Northern
hemisphere at TeV energies, along with Markarian~501 and 1ES2344. In particular, it is the
closest BL Lac object ($\mathrm{z}=0.030$), as well as the first one
discovered in the very high energy (VHE) range: shortly after its detection between
50 MeV and a few GeV by the E{\small GRET} instrument on board the {\it Compton Gamma
Ray Observatory} (Lin, et al. 1992), the Whipple group observed it at a 6.3$\sigma$ level
at TeV energies (Punch, et al. 1992). Since then, it has remained one of the most studied
blazars, and was the object of several multi-wavelength observation campaigns (see, e.g.,
Macomb, et al. 1995, and Takahashi, et al. 1996).\\
Since it started operation in Autumn 1996, the C{\small AT} (Cherenkov Array at Th\'emis) telescope has
been observing Mrk~421 regularly. The next section describes briefly the detector and the analysis method
used to extract the signal. Then, the Mrk~421 data sample and the corresponding light curve up to April 1998
are presented in Sect.~\ref{data.sec}. Finally, the spectral properties observed in 1998 are examined
in Sect.~\ref{sp.sec}, and a discussion is given in Sect.~\ref{discuss.sec}.
\end{figwindow}
\section{The CAT detector: characteristics and analysis method}
\label{detana.sec}
The most complete descriptions of the C{\small AT} experiment and of its specific analysis method can
be found in (Barrau, et al. 1998) and (Le Bohec, et al. 1998), respectively. We just recall
here the few characteristics which are necessary for the understanding of the following sections.
\begin{figwindow}[1,r,
{\mbox{\epsfig{file=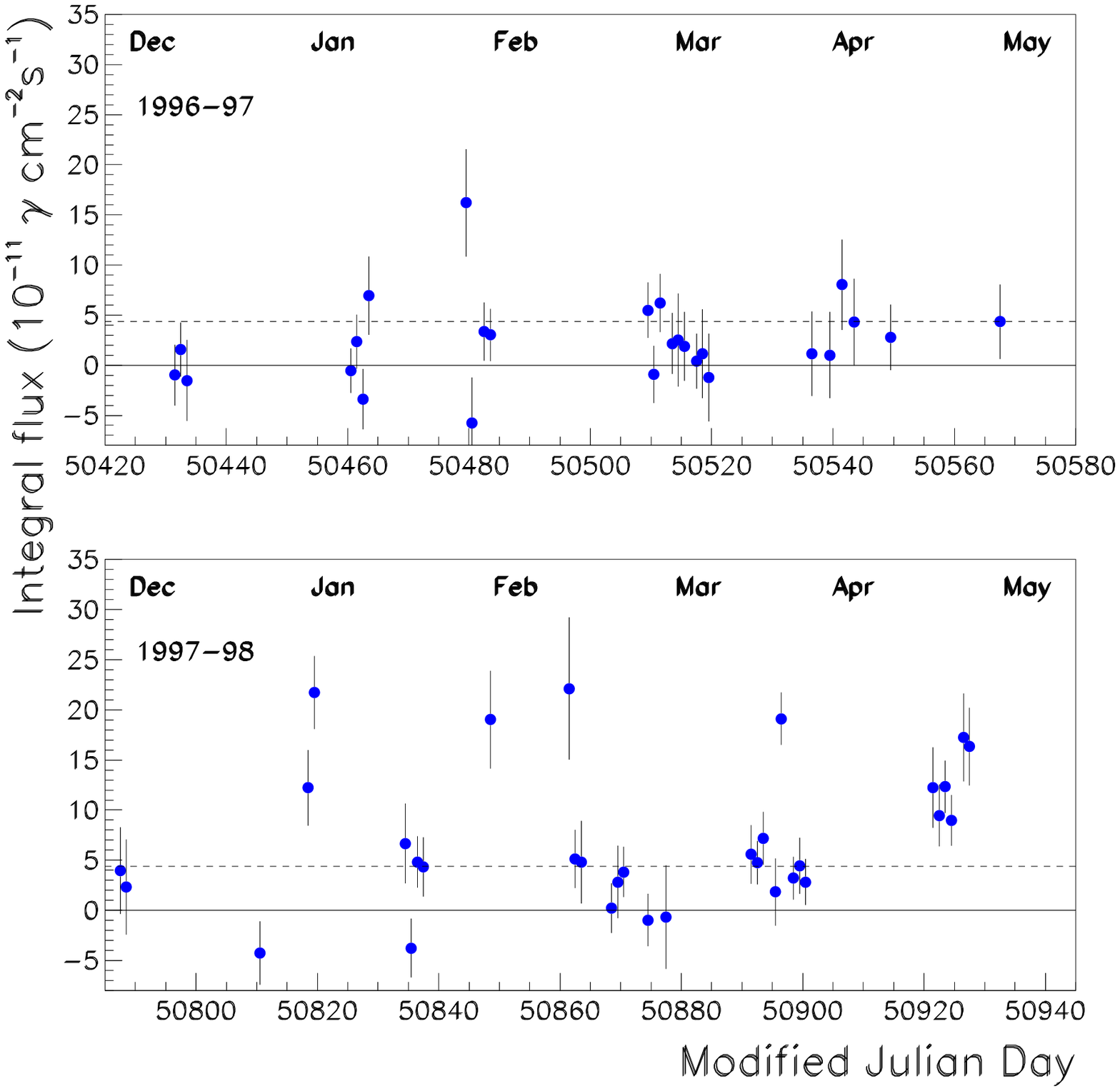,width=4.2in}}},
\label{cl}
{\it Mrk~421 nightly integral flux levels above $250\:\mathrm{GeV}$, for observations
between December 1996 and April 1998. 
The spectral shape as derived in Sect.~\ref{sp.sec} has been assumed to
estimate the integral flux for observations far from the Zenith.
The dashed line represents the mean flux over the two years.}]
Located on the site of the former solar plant Th{\'e}mis (French Pyr\'en\'ees),
the C{\small AT} detector uses the now-validated technique of Cherenkov imaging.
The Cherenkov light emitted by the secondary particles produced
during the development of atmospheric showers is collected
through a large mirror which forms their image in its focal plane. The C{\small AT}
camera, placed $6\:\mathrm{m}$ from its $4.7\:\mathrm{m}$ diameter mirror,
has a $4.8^\circ$ full field of view, which is comprised of a central region of $546$ 
$0.12^{\circ}$ angular diameter phototubes
in a hexagonal matrix and of $54$ surrounding tubes in two ``guard rings''.
Its very high definition, combined with fast electronics, is an efficient solution
of the two main problems one has to face with such $\gamma$-ray detectors: firstly,
the detection threshold, which is determined by the night-sky background, is relatively low
($250\:\mathrm{GeV}$ at Zenith); secondly, the capability of rejecting the enormous
cosmic-ray background (protons, nuclei) is improved by an
accurate analysis based on the comparison of individual images with 
theoretical mean $\gamma$-ray images, i.e. by the fit of an analytical model which
uses the whole information contained in the images' longitudinal light profile.
This fit allows the direction and energy of $\gamma$-ray events to be determined with
good accuracy, with an angular resolution per event of the order of the pixel size,
and an energy resolution about $20$\% (independent of energy) when
selecting showers with an impact parameter less than $130\:\mathrm{m}$.\\
Two main cuts are needed for the $\gamma$-ray selection. After keeping the
highest-energy triggers (total charge $Q_{\mathrm{tot}}>30$ photo-electrons), a first cut
on the probability given by the fit $P({\chi^2})>0.35$ is applied, which selects
the images with a ``$\gamma$-like'' shape. Since $\gamma$-ray images are expected
to point towards the source position in the camera, a second cut $\alpha<6^{\circ}$ is
used, where the pointing angle $\alpha$ is defined as the angle at the image barycentre between 
the actual source position and the reconstructed angular origin of the image.
As a result, this procedure rejects $99.5$\% of hadronic events while keeping $40$\% of
$\gamma$-ray events. A source like the Crab nebula, which is generally considered as the standard
candle for $\gamma$-ray astronomy, can be detected at a $4.5\sigma$ level in one hour and localised
within 1 to 2 arcmin.\\
\end{figwindow}

\section{Data sample}
\label{data.sec}
The $\gamma$-ray emission of Mrk~421 above $250\:\mathrm{GeV}$ has been observed by CAT from December
1996 onwards. Here we consider data taken up to April 1998. A selection based on criteria requiring
clear weather and stable detector operation has been applied. This leaves a total of $73.9$ hours of
on-source ({\small ON}) data, together with $33$ hours on control ({\small OFF}) regions, for a zenith
angle band extending up to $40^{\circ}$. Fig.~\ref{alpha} shows the distribution of the pointing angle
$\alpha$ for the whole sample. With the cuts quoted in Sect.~\ref{detana.sec}, the mean $\gamma$-rate
is found to be of the order of one third of that of the Crab nebula, with a total significance reaching
the $7.6\sigma$ level. After Mrk~501, this makes Mrk~421 the second extragalactic source detected by
C{\small AT}.
\begin{figwindow}[1,r,
{\mbox{\epsfig{file=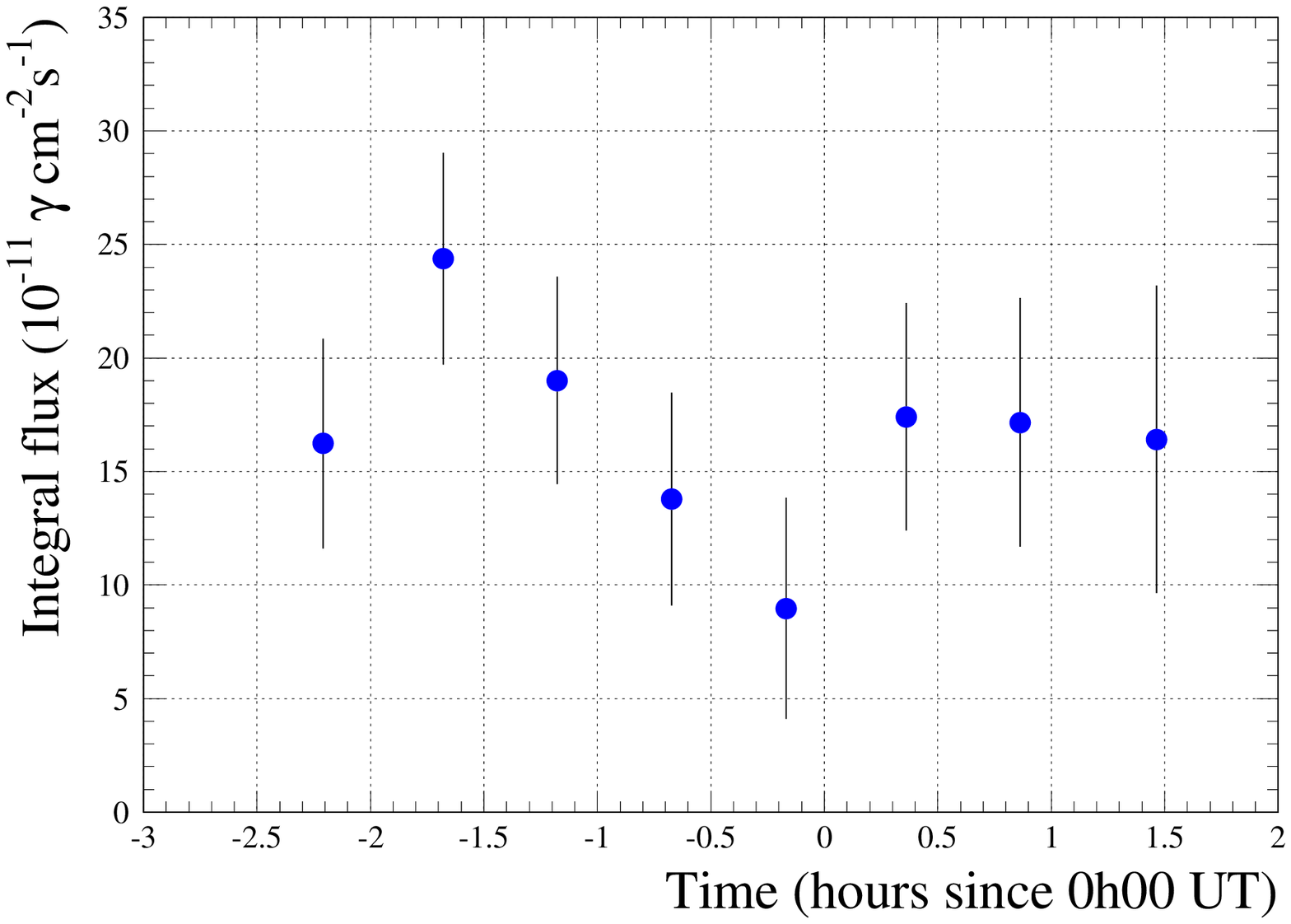,width=3.5in}}},
\label{zoom}
{\it Light curve of the night between 24 and 25 March, 1998 (MJD 50896/50897).
Each point represents a run of $\sim 30$min.}]
As can be seen on Fig.~\ref{cl}, the emission of Mrk~421 changed significantly between
1996-97 and 1997-98: almost quiet in the first period (with a mean integral flux
$\phi(>250\:\mathrm{GeV})= 1.94 \pm 0.65 \times 10^{-11}\:\mathrm{cm^{-2}s^{-1}}$),
the source showed small bursts in the second period, together with a higher mean activity
($\phi(>250\:\mathrm{GeV})= 6.05 \pm 0.54 \times 10^{-11}\:\mathrm{cm^{-2}s^{-1}}$).
Fig.~\ref{zoom} is a zoom on the small flare of the night between 24 and 25 March, 1998,
which reached a maximum of $\sim 2$ times the Crab emission level.
Three days after this flare, a VLBI observation showed a strong decrease of the radio emission
polarization and intensity from the core of the galactic nucleus (Charlot, et al. 1999), in comparison with
a first measurement performed three weeks earlier. The insufficient time-coverage between the radio
and TeV observations unfortunately precludes any further interpretation.
\end{figwindow}

\section{Mrk 421 spectrum in 1998}
\label{sp.sec}
\begin{figwindow}[1,r,
{\mbox{\epsfig{file=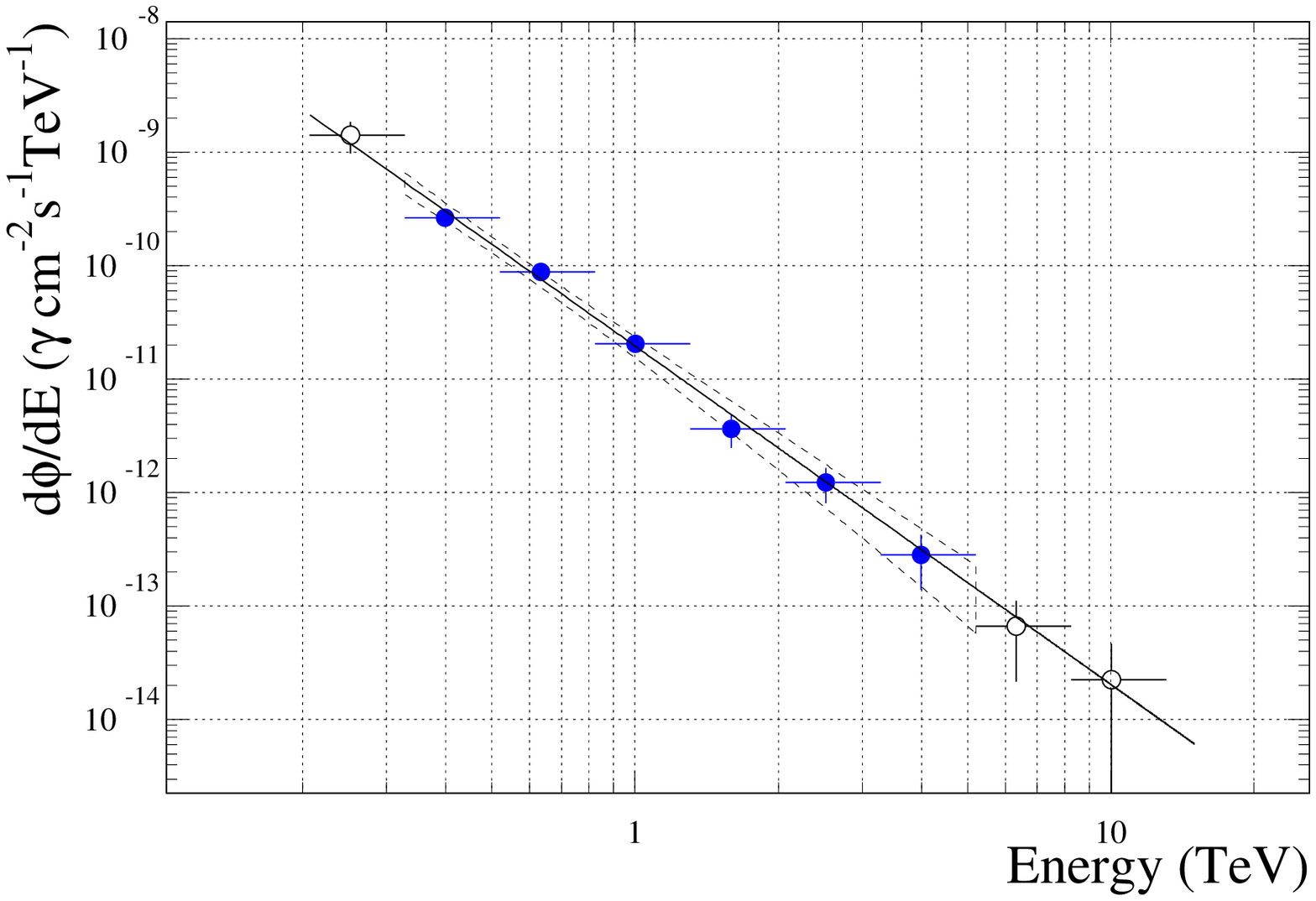,width=4.2in}}},
\label{sp98}
{\it Differential flux of Mrk~421 for the flaring periods in 1998.
Individual intensities per bin are only indicative and the error bars are statistical.
The full line represents the power law parameters estimated by a likelihood method,
in which only the fullfilled circles were used (see text).
This method gives also the $2\sigma$ error ``box'', whose shape reflects the correlation
between the parameters. This latter vanishes at the decorrelation energy
$\mathrm{E_d}=580\:\mathrm{GeV}$.}]
A spectrum was derived for the flaring periods of Mrk~421 in 1998. The data used in this
section have been further limited to zenith angles $<16^{\circ}$, i.e. to a configuration for which
the detector calibration has been fully completed. By minimizing systematic effects, this allows a robust
spectral determination. Another favourable factor is the low night-sky background in the field of view,
due to the lack of bright stars around the source.\\
The spectrum was derived in two steps: firstly, the number of events for each energy and zenith angle
bin within the cuts quoted in Sect.~\ref{detana.sec} was determined for {\small ON} and {\small OFF} runs;
secondly, taking into account the trigger and cut efficiencies as well as the energy resolution of the
telescope, a maximum likelihood estimation of the spectral parameters has been applied for the simple power
law hypothesis ($\mathcal{H}_0$)
$\mathrm{d}\Phi/\mathrm{d}E_\mathrm{TeV}=\phi_0 E_\mathrm{TeV}^{-\gamma}$.
As can be seen in Fig.~\ref{sp98}, this accounts very well for the observed spectrum between
$330\:\mathrm{GeV}$ and $5.2\:\mathrm{TeV}$, which was the energy range considered.
With a correlation coefficient of $-0.68$, the fitted parameters are:\\
$\phi_0 = \mathrm{ 1.96 \pm 0.20^{stat} \pm 0.12^{sys-MC} }
\stackrel{+1.07}{_{-0.59}}^\mathrm{sys-atm}
\times 10^{-11}\:\mathrm{cm^{-2}s^{-1}TeV^{-1}}$ and \\
$\gamma = 2.96 \pm \mathrm{ 0.13^{stat} \pm 0.05^{sys-MC}}$.\\
Systematics errors due to limited Monte Carlo statistics (noted ``sys--MC'') in the determination of
the effectice detection area affect both $\phi_0$ and $\gamma$. The second systematics errors
(noted ``sys--atm'') come from the uncertainty on the absolute energy scale. They are due to
variations of the transparency of the atmosphere and affect only the absolute flux.
\end{figwindow}

\section{Discussion}
\label{discuss.sec}
The spectral shape of Mrk~421 observed by C{\small AT} in 1998 is fully compatible with
recent results from the H{\small EGRA} collaboration (Aharonian, et al. 1999).
It is however in contrast with the former behaviour
of the source at the time of the 1995 and 1996 flaring periods, as observed by the Whipple
group, who found $\gamma = 2.54 \pm \mathrm{0.03^{stat} \pm 0.10^{sys}}$ (Krennrich, et al.
1999). The higher value of the differential spectral index found here thus seems to
favour a correlation between the intensity level and the spectral shape. Actually, such a
correlation has been recently reported by C{\small AT} on the extreme blazar
Mrk~501 (Djannati-Ata\"{\i}, et al. 1999; see also Tavernet, et al. 1999).
Since the spectrum of the latter is now known to be curved for very intense
emissions, the hypothesis ($\mathcal{H}_1$) of a curved shape 
$\mathrm{d}\Phi/\mathrm{d}E_\mathrm{TeV}=\phi_0 E_\mathrm{TeV}^{-(\gamma +
\beta\log_{10}\!E_\mathrm{TeV})}$ was also considered for Mrk~421.
The fitted curvature term is $\beta=0.28 \pm 0.49$, compatible with zero. The likelihood ratio
$\lambda = -2 \times \log \frac{\mathcal{L}(\mathcal{H}_0)}{\mathcal{L}(\mathcal{H}_1)}$,
which gives an estimate of the relevance of $\mathcal{H}_1$ with respect to
$\mathcal{H}_0$ and behaves asymptotically like a $\chi ^2$ with one d.o.f.,
is $0.34$, corresponding to a chance probability of $0.56$. This confirms the absence of any obvious
spectral curvature for Mrk~421, as already indicated by Krennrich, et al. 1999.
In the framework of leptonic models (Ghisellini, et al 1998), which succesfully explain the
Mrk~501 spectral energy distribution in the X-ray and VHE $\gamma$-ray ranges
(Djannati-Ata\"{\i}, et al. 1999),
the present result implies that the peak energy of the inverse Compton contribution of Mrk~421 is
significantly lower than the C{\small AT} detection threshold.
This is not surprising since the corresponding synchrotron peak is lower than that of Mrk~501, and
since leptonic models predict a strong correlation between X-rays and $\gamma$-rays. Moreover,
such a correlation was directly observed on Mrk~421 in Spring 1998, during a coordinated
observation campaign involving ground-based Cherenkov imaging telescopes (Whipple, H{\small EGRA},
and C{\small AT}) and the A{\small SCA} X-ray satellite (Takahashi, et al. 1999).


\vspace{1ex}
\begin{center}
{\Large\bf References}
\end{center}
%
Aharonian, F.A., et al. 1999, submitted to A\&A\\
Barrau, A., et al. 1998, Nucl. Instr. Meth. A416, 278\\
Charlot, P., et al. 1999, Proc. $19^{th}$ Texas Symposium, Paris 1998, in press\\
Djannati-Ata\"{\i}, A., et al. 1999, submitted to A\&A\\
Ghisellini, G., et al. 1998, MNRAS 301, 451\\
Krennrich, F., et al. 1996, ApJ 481, 758\\
Krennrich, F., et al. 1999, ApJ 511, 149\\
Le Bohec, S., et al. 1998, Nucl. Instr. Meth. A416, 425\\
Lin, Y.C., et al. 1992, ApJ 401, L61\\
Macomb, D.J., et al. 1995, ApJ 449, L99\\
Punch, M., et al. 1992, Nature 358, 477\\
Takahashi, T., et al. 1996, ApJ 470, L89\\
Takahashi, T., et al. 1999, to be published\\
Tavernet, J.-P., et al. 1999, this Proc., OG 2.1.08\\

\end{document}